
%
%
%

\def\predefine#1#2{\let#1=#2}
\font\twelverm=cmr12
\font\twelvei=cmmi12
\font\twelveit=cmti12
\font\twelvesl=cmsl12
\font\twelvesy=cmsy10 at 12pt
\font\twelvebf=cmbx12
\font\twelvett=cmtt12
\def\twelvepoint{\def\rm{\fam0\twelverm}%
\textfont0=\twelverm \scriptfont0=\tenrm \scriptscriptfont0=\sevenrm
\textfont1=\twelvei \scriptfont1=\teni \scriptscriptfont1=\seveni
\textfont2=\twelvesy \scriptfont2=\tensy \scriptscriptfont2=\sevensy
\textfont3=\tenex \scriptfont3=\tenex \scriptscriptfont3=\tenex
\def\it{\fam\itfam\twelveit}%
\textfont\itfam=\twelveit
\def\sl{\fam\slfam\twelvesl}%
\textfont\slfam=\twelvesl
\def\bf{\fam\bffam\twelvebf}%
\textfont\bffam=\twelvebf
 \scriptfont\bffam=\tenbf \scriptscriptfont\bffam=\tenbf
\def\tt{\fam\ttfam\twelvett}
\textfont\ttfam=\twelvett
\def\smc{\twelvesmc}
\baselineskip 14pt%
\abovedisplayskip 14pt plus 3pt minus 10pt%
\belowdisplayskip 14pt plus 3pt minus 10pt%
\abovedisplayshortskip 0pt plus 3pt%
\belowdisplayshortskip 8pt plus 3pt minus 5pt%
\parskip 3pt plus 1.5pt
\setbox\strutbox=\hbox{\vrule height10pt depth4pt width0pt}%
\rm}
\output{\plainoutput}
\predefine\oldtenpoint{\tenpoint}
\redefine\tenpoint{\twelvepoint}
\tenpoint
%
%
\voffset=-0.5cm
\hsize=6in
\vsize=8.5in
%
%
\def\nn{{\ | \!\!\! |\ }}

\def\vek#1{{\vec #1}}
\def\real{{I\!\! R}}
\def\Tr{{\rm Tr}}
\def\ad{{\rm ad}}
\def\half{{\scriptstyle{1\over 2}}}
\def\halfi{{\scriptstyle{i\over 2}}}
\def\twothird{{\scriptstyle{2\over 3}}}
\def\cO{{\cal O}}
%
%
%
\line{\hbox{\hskip13.0cm THU-92/20}}
\line{\hbox{\hskip12.1cm CMU-HEP92-19}}
\line{\hbox{\hskip12.9cm August 1992\hfil}}
\vskip.3cm
%
%
\centerline{\bf NON-PERTURBATIVE ANALYSIS, GRIBOV HORIZONS AND }
\centerline{{\bf THE BOUNDARY OF THE FUNDAMENTAL DOMAIN}
\footnote\dag{\tenrm Contributed talk presented at 21st Conference
on Differential Geometric Methods in Theoretical $\hphantom{AB}$
Physics (XXI DGM), June 5-9, Nankai University, Tianjin, P.R. China.}}
%
\vskip.7cm
\centerline{PIERRE VAN BAAL \parindent=5mm\footnote*{KNAW fellow}}
\centerline{\it Institute for Theoretical Physics, P.O.Box 80.006,}
\centerline{\it NL-3508 TA Utrecht, The Netherlands}
%
\vskip.3cm
\centerline{and}
%
\vskip.3cm
\centerline{R.E. CUTKOSKY}
\centerline{\it Physics Department, Carnegie Mellon University,}
\centerline{\it Pittsburgh, PA 15213, USA}
%
\vskip.6cm
\centerline{ABSTRACT}
{\narrower
\baselineskip=12pt{\tenrm
In this contribution to the proceedings we will describe some of the
details for constructing the Gribov horizon and the boundary of the
fundamental modular domain, when restricting to some low energy modes
of pure SU(2) gauge theory in a spherical spatial geometry.
The fundamental domain is a one-to-one
representation of the set of gauge invariant degrees of freedom, in
terms of transverse gauge fields. Boundary identifications are the
only remnants of the Gribov copies.}
\par}
\vskip.5cm
\baselineskip=14pt
\noindent{\bf 1. Introduction}

At the conference one of us
gave an overview of the finite volume analysis on a torus,
relevant for comparison with lattice Monte Carlo results~[1].
Recently, for technical reasons, this analysis was extended to a
spherical spatial geometry~[2], results of which were only briefly
touched upon during the talk. The other author has, from a different
perspective, been interested in this geometry for quite some years
now~[3]. Thus we found some common interest, whose hitherto unpublished
fruits we will discuss in these proceedings. It involves computing the
Gribov horizon and the boundary of the fundamental domain in a
spherical spatial geometry. The remaining part of the talk is summarized
in the contribution to the Shanxi conference~[4].
For the applications and further references see~[1-4].

In gauge theories one can fix to the transverse or Coulomb gauge,
$\partial_iA_i=0$, using a functional method, which allows one to
pick from the different Gribov copies~[5] (transverse gauge fields
that are nevertheless gauge equivalent) an (almost) unique
representative. The collection of these configurations should thus
form a fundamental modular domain, in other words it should form
a one-to-one mapping with the Yang-Mills configuration space.
The relevant functional is the $L^2$-norm of the gauge field:
$\nn A_i\nn^2=\int_M\Tr(A_i^\dagger A_i)$.
For each gauge invariant field configuration this gives a Morse
functional on the gauge orbit. One easily verifies that stationary
points of this Morse functional satisfy the Coulomb gauge condition
and that the Hessian (second order derivative) at the stationary
point is precisely given by the Faddeev-Popov operator, whose
determinant measures the volume of the gauge orbit. It is natural
to choose among the various Gribov copies the one with lowest
norm. The collection of transverse fields thus obtained has
a boundary. Points on this boundary are usually degenerate in norm and
gauge equivalent to at least one other point on the boundary~[6]. This
gauge equivalence induces boundary identifications that make the
set into a fundamental modular domain $\Lambda$ (no Gribov copies occur
at the interior). In general $\Lambda$ is well contained in the Gribov
region $\Omega$, which is by definition the collection of transverse
potentials for which the norm functional (when considered as a function
of the gauge orbit) has a local minimum, i.e. the Hessian or Faddeev-Popov
operator $FP(A)=-\partial_iD_i(A)$ is positive semi-definite
($D_i(A)$ is the
usual adjoint covariant derivative). There can~[6], however, be points at
the boundary of $\Lambda$ that coincide with the boundary of $\Omega$.
Since the latter is the Gribov horizon, which is where the lowest
eigenvalue of the Faddeev-Popov operator (and hence the Faddeev-Popov
determinant) vanishes, these points will still require some extra
care. They have, so far, not been considered in the subsector of the
theory we will be studying here.

\noindent{\bf 2. Defining the subspace}

The subsector we will study is given by
18 modes that are degenerate in energy, to lowest (quadratic) order.
There are only twelve modes with a lower energy, also all degenerate.
These 18 modes contain degrees of freedom  relevant for the tunnelling
from the $A=0$ vacuum to the two vacua that have Chern-Simons
number $Q(A)={1\over 8\pi^2}\int_{S^3}\Tr(A\wedge dA+\twothird A\wedge
A\wedge A)$ one or minus one.
Here $A=A_idx_i=iA^a_i\tau_adx_i/2$ is the connection one-form
for a SU(2) vector potential $A_i^a$ on the three-sphere.
In other words the
two vacua with $Q(A)=\pm1$, are transverse vector potentials, that
are pure gauge ($A=gdg^{-1}$) with a gauge function $g$
that has a winding number $n(g)=\pm1$, where $n(g)={1\over 24\pi^2}
\int_{S^3}\Tr((g^{-1}dg)^3).$ The tunnelling between
the degenerate vacua is of course described by instantons, easily obtained
from the well-known instanton solutions on
$\real^4$ by the conformal transformation that relates
$S^3\times\real$ to $\real^4$. To be precise, the
18 modes contain the tunnelling paths
that describe the transition over the lowest barrier separating the
nearest-neighbour vacua. This particular saddle point of the energy
functional is also known as a sphaleron.

One way of describing
these modes is by using~[2] the 't Hooft $\eta$ symbols
(useful in giving an explicit expression for the instanton vector
potentials), to define a framing of the three-sphere.
It is easy to specify the $\eta$ symbol in terms of a basis
(and its dual) of the unit quaternions $\sigma_\mu$ ($\bar\sigma_\mu$)
with $\sigma_4=\bar\sigma_4=1$ and $\sigma_a=-\bar\sigma_a=i\tau_a$.
One has $2i\eta_{\mu\nu}^a\tau_a=\sigma_\mu\bar\sigma_\nu-
\sigma_\nu\bar\sigma_\mu$ and if we parametrize the three-sphere
by the unit vectors $n_\mu$ in four dimensions, one can define a
dreibein by $e_\mu^i=\eta_{\mu\nu}^in_\nu$. The 18 modes now
split in two categories: there are 9 modes described by constant
components $A_i^a(n)=c_i^a$ (it is essential that the index $i$ refers
to a flat index with respect to the above defined framing), whereas
the other 9 modes are constant up to a coordinate-dependent rotation,
$V_i^b(n)=\half\Tr(n^\mu\sigma_\mu\sigma_in^\nu\bar\sigma_\nu\sigma_b)$,
of the dreibein, i.e. $A_i^a(n)=-V_i^b(n)d_b^a$.
One easily verifies that both vector potentials are transverse.
It is furthermore not too difficult to show that (minus) the rotated frame
is precisely the one obtained by replacing $\eta$ by $\bar\eta$
(defined by $2i\bar\eta_{\mu\nu}^a\tau_a=\bar\sigma_\mu\sigma_\nu-
\bar\sigma_\nu\sigma_\mu$), which is used to express the
anti-instantons. The instanton is represented by
$c_i^a=-2\delta_i^a/(1+e^{-2t})$ and $A_0=0$,
with an identical expression for the anti-instanton in terms of
$d_i^a$. In these proceedings we will restrict ourselves
to the two-dimensional cross-section of the field space corresponding to
the direction of these particular instanton and anti-instanton
configurations which, since they describe tunnelling through the
sphaleron (the lowest barrier separating two nearest-neighbour vacua),
were called sphaleron modes~[2]. Thus we will consider
$A_i^a(n)=vV_i^a(n)-u\delta_i^a$.

The classical
vacua in the $(u,v)$ plane are located at $(0,0)$, $(0,2)$ and $(2,0)$,
whereas the sphalerons can be found at $(0,1)$ and $(1,0)$.
Furthermore, applying the gauge transformation $g(n)=n^\mu\bar\sigma_\mu$
(which has winding number $n(g)=-1$)
to the configuration $(u,0)$ can be shown~[2] to yield $(0,2-u)$. As
it should, this maps the vacua $(2,0)$ and $(0,0)$ (which have
respectively $Q(A)=1$ and $Q(A)=0$) to the vacua $(0,0)$ and $(0,2)$
(where $(0,2)$ has $Q(A)=-1$). Furthermore it maps the sphaleron at
$(1,0)$ (with $Q(A)=\half$) to the sphaleron at $(0,1)$
(with $Q(A)=-\half$).

\noindent{\bf 3. The Gribov horizon}

To diagonalize the Faddeev-Popov operator $FP(A)$ in the subspace
$(u,v)$ it is convenient to introduce angular momentum operators~[2]
$L_1^a=\halfi e^a_\mu\partial_\mu=-\halfi\eta_{\mu\nu}^an^\mu\partial_\nu$,
$L_2^a=-i\half\bar\eta_{\mu\nu}^an^\mu\partial_\nu$,
$T^a=\half\ad(\tau_a)$, and $\vek J=\vek L_1+\vek L_2+\vek T$. It is not
too difficult to check that $FP(u,v)$ commutes both with
$\vek L_1^2=\vek L_2^2$  and with $\vek J$. For this it is convenient
to write $FP(u,v)=4\vek L_1^2+2u\vek T\cdot\vek L_1-v~\ad(\bar\sigma_\mu
n^\mu\vek\tau\sigma_\nu n^\nu)\cdot\vek L_1$; only the last term in this
expression requires some care in computing the commutators with
$\vek J$ and $\vek L_1^2$. Thus $FP(u,v)$ can be diagonalized in
the subspace defined by $\ell\neq 0$, $j$ and $j_z$. There is an obvious
degeneracy in $j_z$ and the eigenvalues will be denoted by
$\lambda_{2\ell,j}(u,v)$. Note that
the Coulomb gauge does not fix the constant gauge transformations, which
means that the constant modes ($\ell=0$) are eliminated from the
spectrum of $FP(A)$. The remaining invariance under constant gauge
transformations is easily taken into account without further gauge fixing.
As the Gribov horizon is defined as the set of configurations where
the lowest eigenvalue of $FP(A)$ vanishes, it suffices to diagonalize
$FP(u,v)$ for $\ell=\half$. There are 12 eigenfunctions in this sector
that split in one $j=0$ singlet $(n_a\tau_a)$, two $j=1$ triplets
($n_0\tau_a$ and $\varepsilon_{abc}n_b\tau_c$) and one $j=2$ quintet
($n_a\tau_b+n_b\tau_a-\twothird n_c\tau_c\delta_{ab}$). For the singlet
and quintet the problem of diagonalizing $FP(u,v)$ becomes
one-dimensional, whereas for the two triplets one has to diagonalize
a $2\times 2$ matrix. One easily checks that $\lambda_{1,0}=3-2s$,
$\prod\lambda_{1,1}=(9-3s-2p^2)^3$ and $\lambda_{1,2}=3+s$
(for convenience we introduced the scalar and pseudoscalar (even and odd)
helicity combinations $s\equiv u+v$ and $p\equiv u-v$).
In fig.~1 the full curves give the solutions of $\lambda_{1,j}(u,v)=0$,
whereas the Gribov horizon is indicated by fat sections. The result
is in accordance with the convexity of $\Omega$. This convexity is a
simple consequence of the linear dependence of $FP(A)$ on the vector
potential. Note that the sphaleron configurations are well within
the Gribov region, but that the vacua nearest to $A=0$ are outside (but
not on) the Gribov horizon. Nevertheless, it is easily proved~[6] that at
these vacua the Faddeev-Popov determinant has to vanish. It turns
out that the relevant eigenvalues (there are actually 3 of them) that
vanish at the vacua nearest to $A=0$ have $\ell=1$. So it is instructive
to diagonalize $FP(u,v)$ in this sector too. One finds one $j=0$
singlet, three $j=1$ triplets, two $j=2$ quintets and one $j=3$ septet.
We thus have to diagonalize at most a $3\times 3$ matrix to obtain the
eigenvalues for $FP(u,v)$, with the following result $\lambda_{2,0}=8-2s$,
$\prod\lambda_{2,1}=512((8-2s)^2-s^2+(2s-7)p^2)^3$,
$\prod\lambda_{2,2}=(64-s^2-3p^2)^5$ and $\lambda_{2,3}=8+2s$.
This allows us to determine the location of the zero's for the Faddeev-Popov
determinant in this sector, indicated in fig.~1 by the dashed curves.
%
%
\vskip13cm\vfill
{\narrower\baselineskip=12pt{\tenrm\noindent Figure 1: Location of the
classical vacua (large dots), sphalerons (smaller dots), zeros of the
Faddeev-Popov determinant (full curves at $\ell=\half$, dashed curves
at $\ell=1$), the Gribov horizon (fat sections) and part of the boundary
of the fundamental domain (dotted curves) in the plane specified
by $A_i^a=vV_i^a(n)-u\delta_i^a$.}\par}

\noindent{\bf 4. The boundary of the fundamental modular domain}

To construct the boundary of the fundamental modular domain requires
one to find transverse gauge copies that are degenerate in norm.
This is unfortunately a rather non-local problem. We can,
however, make use of the fact that the two sphalerons are transverse,
have the same norm and are gauge copies of each other, as we discussed
above. We can therefore start from $(0,1)$ and construct $(u,r(u))$,
expanding in powers of $u$, such that the gauge transformation of
$A(u,r(u))$ (coinciding with the sphaleron configuration at $(1,0)$)
is again transverse and has the same norm
(note that for $u\neq0$, the gauge transformed potential is {\it
not} part of the $(u,v)$ plane).
Obviously, also $(r(v),v)$ is part of the boundary of the fundamental
modular domain. These two branches are indicated by the dotted curves
in figure 1. To find the desired gauge transformation one
writes $g(n)=n^\mu\sigma_\mu\exp(X(n))$ and one minimizes
$M(u,v)\equiv\nn g(n)\vek A(u,v) g(n)^{-1}+g(n)\vek\partial
g(n)^{-1}\nn^2-\nn\vek A(u,v)\nn^2$ with respect to
$X$. It can be verified a posteriori (by checking transversality of the
gauge transformed vector potential) that the solution is of the form
$X(n)=if(n_0)\vek n\cdot\vek\tau$. Using this as an
ansatz, reduces the problem of finding the stationary solution to
solving an ordinary 2nd order differential equation. Its
solution can be written as
$f(x)=x\sum_{j=1}\sum_{k=0}^{j-1}a_{j,k}(v)u^jx^{2k}$. One finds:
$a_{1,0}(v)=2(2+v)^{-1}$, $a_{2,0}=-2(v^2+6v-16)(2+v)^{-3}
(10+v)^{-1}$, $a_{2,1}(v)=4(6+v)(v+2)^{-2}(v+10)^{-1}$, etc.
By solving $M(u,r(u))=0$ we find
$$r(u)=1-{1\over9}u^2-{2\over81}u^3-{25\over2673}u^4-{1238\over264627}u^5
-{172442\over66950631}u^6-{687429956\over457339760361}u^7+\cO(u^8).$$
The result is exhibited in figure 1 by the dotted curves.
Similar to the Gribov region, the fundamental modular domain is convex and
its boundary has ``corners'', where different branches of copies that
are degenerate in norm intersect. Usually this will keep
$\partial\Lambda$ from touching the Gribov horizon~[6].

\noindent{\bf Acknowledgements}
This research has been made possible by a fellowship of the Royal
Netherlands Academy of Arts and Sciences and by the U.S. Dept. of Energy
under grant No. DE-FG02-91ER40682.

\noindent{\bf References}
\item{1.} J. Koller and P. van Baal, {\it Nucl. Phys.} {\bf B302} (1988)
1; P. van Baal, {\it Phys. Lett.} {\bf B224} (1989) 397.
\item{2.} P. van Baal and N.D. Hari Dass, ``The theta dependence
beyond steepest descent'', to appear in {\it Nucl. Phys.} {\bf B},
Utrecht preprint THU-92/03, January 1992.
\item{3.} R.E. Cutkosky, {\it J. Math. Phys.} {\bf 25} (1984) 939;
R.E. Cutkosky and K. Wang, {\it Phys. Rev.} {\bf D37} (1988)
3024; R.E. Cutkosky, {\it Czech. J. Phys.} {\bf 40} (1990) 252.
\item{4.} P. van Baal, ``Topology of Yang-Mills configuration space'',
Utrecht preprint THU-92/15, to appear in the proceedings of ISATQP,
June 11-16, Shanxi University, (Science Press of China).
\item{5.} V. Gribov, {\it Nucl. Phys.} {\bf B139} (1978) 1.
\item{6.} P. van Baal, {\it Nucl. Phys.} {\bf B369} (1992) 259.
\bye